\documentstyle[psfig]{caosp}
\begin{document}
\pubyear{1993}
\volume{23}
\firstpage{7}
\htitle{Binaries among CP stars}
\hauthor{P. North {\it et al.}}
\title{Binaries among Ap and Am stars}
\author{P. North\inst{1} \and N. Ginestet\inst{2} \and J.-M. Carquillat\inst{2}
\and F. Carrier\inst{3} \and S. Udry\inst{3}}
\institute{Institut d'Astronomie de l'Universit\'e de Lausanne,\\
CH-1290 Chavannes-des-Bois, Switzerland \and
 Observatoire Midi Pyr\'en\'ees, 14 avenue Edouard Belin, F-31400 Toulouse,
 France
\and Observatoire de Gen\`eve, CH-1290 Sauverny, Switzerland}

\date{\today}
\maketitle
\begin{abstract}
The results of long-term surveys of radial velocities of cool Ap and Am
stars are presented. There are two samples, one of about 100 Ap stars 
and the other of 86 Am stars. Both have been observed with the CORAVEL scanner
from Observatoire de Haute-Provence (CNRS), France.

The conspicuous lack of short-period binaries among cool Ap stars seems
confirmed, although this may be the result of an observational bias;
one system has a period as short as 1.6 days. A dozen new orbits could be
determined, including that of one SB2 system. Considering the mass functions
of 68 binaries from the literature and from our work, we conclude that
the distribution of the mass ratios is the same for the Bp-Ap stars than
for normal G dwarfs.

Among the Am stars, we found 52 binaries, i.e. 60\%; an orbit could be
computed for 29 of them. Among these 29, there are 7 SB2 systems, one triple
and one quadruple system. The 21 stars with an apparently constant radial
velocity may show up later as long-period binaries with a high eccentricity.
The mass functions of the SB1 systems are compatible with cool main-sequence
companions, also suggested by ongoing spectral observations.
\end{abstract}

\section{Introduction}
The question here is that of the possible influence of a binary companion upon
chemical peculiarity: does an Ap or Am star ``need'' a companion, or conversely,
does it need to be single? If there is a companion, what is its nature?
Finally, fundamental parameters like mass, radius and age are better
constrained for the components of binary stars than for isolated ones.

The first systematic search for spectroscopic binaries among Bp-Ap stars was
made by Abt \& Snowden (1973) on a sample of 62 northern bright stars. They
found a rate of binaries of only 20 percent among the magnetic Ap stars and of
43 percent among HgMn stars (the rate among normal stars is $47\pm 5$ percent
among normal B0 to M stars according to Jaschek \& Gomez 1970). Aikman (1976)
studied 80 HgMn stars (instead of 15 for Abt \& Snowden)
and confirmed a normal rate
with 49 percent of binaries. The most recent and complete review is that of
Gerbaldi et al. (1985), while North (1994) published some new results on cool
Ap stars obtained with the CORAVEL spectrovelocimeter.

On the basis of a sample of 25 Am stars, Abt (1961) concluded
that all are binaries, while Abt \& Levy (1985) confirmed the
high rate of binaries in a larger sample but also found some stars to be 
apparently single. For both Ap and Am stars, only bright objects had been 
examined, so it was judged worthwhile to extend the survey to fainter and more 
numerous ones.

This contribution is kind of a progress report on a survey of cool Ap and
Am stars made with the CORAVEL scanner in the northern hemisphere.

\section{Samples and observations}
All observations have been made at Observatoire de Haute-Provence (CNRS),
France, with the CORAVEL spectrovelocimeter (Baranne et al. 1979) attached to
the 1m Swiss telescope. The CORAVEL scanner is optimized for cool stars, since
the mask yielding the correlation function was made from the spectrum of the
yellow giant Arcturus. Nevertheless, even stars with as hot a type as A0 may
provide an autocorrelation dip, provided they rotate slowly enough and have
enhanced metallic lines, which is precisely the case of many (though not all)
Ap and Am stars.

\subsection{Cool (SrCrEu) Ap stars}
Some well-known magnetic Ap stars had been already measured by Dr. Michel
Mayor and coworkers as early as 1980, but a more complete sample has been
defined in 1985 from the following criteria:\\
-- Ap classification from e.g. the compilation of Bertaud \& Floquet (1974)\\
-- Ap stars having Geneva colours\\
-- $T_{\rm eff} \la 10000$~K according to Geneva photometry\\
-- declination $\delta \geq -20\degr$\\
-- $V\leq 8.6$ (some fainter stars included as well)\\
This makes a total of 160 stars, but 48 of them proved to yield no
autocorrelation dip because of fast rotation. Indeed, the dip becomes too flat
when $v\sin i \ga 30$~km\,s$^{-1}$, so that our sample is severely biased
against high $v\sin i$. This implies that the sample is also biased against
periods shorter than about 4 days, if synchronism is assumed and if
$i\sim 90\degr$ (assuming also the orbital and equatorial planes to be 
coplanar); systems with shorter periods may of course be observed,
provided the inclination
$i$ is small enough. Among the 112 stars left, 13 are more probably Am than Ap
stars because some early, unreliable classification sources were used, and a
few more had a so wide and shallow correlation dip that they were observed only
once or twice. Therefore, we are left with 95 objects measured $\geq 3$ times
(typically 6-8 times) over a time span of 12 to 17 years.
\subsection{Am stars}
Systematic observations aimed at studying the binaries began in 1992.
The sample was defined as including\\
-- northern objects\\
-- metallic type F2 or later according to the compilation by Hauck (1986);
some objects from Bidelman (1988) and from Hynek (1938) were also included\\
-- spectroscopic binaries with known orbits were excluded from the sample\\
As for the sample of Ap stars, this one is biased against high $v\sin i$
(therefore against very short orbital periods). But it is also biased, to some
extent, against large $V_r$ amplitude, short period spectroscopic binaries,
because objects with known and published orbits -- which are more easily 
discovered when the period is short -- are excluded. There are 230 objects in 
this sample, among which 158 have been observed to date. However, as many as 72 
stars did not yield any autocorrelation dip, probably because of their high 
rotational velocity, so we are left with 86 objects.

\section{Results}
\subsection{Ap stars}
Because of the inhomogeneous distribution of the elemental abundances on the
surface of these stars, many of them are spectrum variables and present radial
velocity variations due to rotation and abundance patches. This problem was
avoided by Abt \& Snowden (1973) by using hydrogen rather than metallic lines,
but this is of course not possible with CORAVEL. We are forced to identify these
stars {\it a posteriori}, which is possible at least in part, thanks to some 
typical characteristics like wide and shallow correlation dip, short timescale
(a few days) and relatively low-amplitude variability ($\sim 20$~km\,s$^{-1}$)
leading to a very small ``mass function'' if one tries to fit an orbit.
Taking this into account, one has 2 SB2 and 24 certain SB1 binaries among all
95 stars, which makes 27 percent. This percentage is smaller than that of normal
stars ($47\pm 5$~\%). 

What is more interesting is that we found several
long-period binaries but relatively few short period ones, which seems to
confirm the lack of tight orbits among Ap stars in spite of the bias we have
against periods shorter than 3-4 days. This is entirely consistent with the
results of Gerbaldi et al. (1985), who found no orbital period shorter than
about 3 days. However, we have found HD 200405 to be a spectroscopic binary
with $P_{\rm orb} = 1.635$~days, which is the shortest orbital period known
among Ap stars. Figure 1a shows the $e$ vs. $\log P$ diagram for all known
spectroscopic binaries with an Ap stars, including 9 new orbits determined with
CORAVEL. This diagram resembles very closely that of normal stars (see Fig. 1
of Gerbaldi et al. 1985), the only difference being the cutoff at about 0.5.
This cutoff suffices to explain the peculiar distribution of eccentricities
of the Ap binaries shown in Fig. 7 of Gerbaldi et al. (1985): the lack of low 
eccentricities is linked with the lack of very short periods among Si and SrCrEu
binaries; HgMn stars tend to have shorter periods than the latter,
which also explains why there
are more circular orbits among them. Thus, the distribution of eccentricities
does not appear to be peculiar in itself but only reflects the distribution
of orbital periods.
\vspace{-0.5cm}
\begin{figure}[hbt]
\psfig{figure=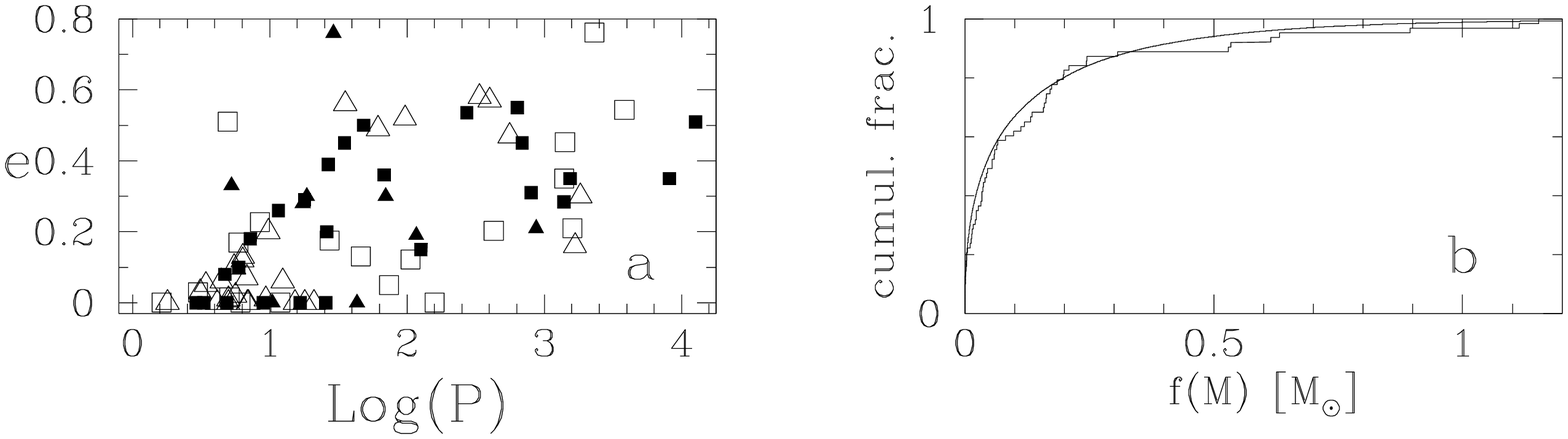,height=9.5cm}
\vspace{-5cm}
\caption{{\bf a:}$e$--$\log P$ diagram for Ap stars. Key to symbols: open 
triangles=HgMn, black triangles=He~weak, open squares=SrCrEu, black squares=Si.
{\bf b:} Cumulative distribution of all Ap binaries (including HgMn) compared
with a simulated one.}
\end{figure}
Figure 1b shows the cumulative distribution of the mass functions of all 
binaries hosting an Ap star (including the HgMn, but the result would be the
same without them). The 
continuous line is the result of a Monte-Carlo simulation where the mass ratio 
is a gaussian, as proposed by Duquennoy \& Mayor (1991) for the nearby G dwarfs;
the mass function of the primaries is Salpeter's one multiplied by the relative 
mass function of Ap stars given by North (1994). The observed
and simulated distributions are in perfect agreement, suggesting that the
unseen companion of binary Ap stars has nothing exotic; in particular, it is
generally not a white dwarf.
\subsection{Am stars}
We have found as yet 52 binaries in our sample of 86 objects, i.e. 60 percent.
29 orbits could be determined for 19 SB1 systems, 7 SB2 ones, one triple and
one quadruple system. 13 stars have not been measured often enough to conclude
about their possible $V_r$ variability, but the remaining 21 seem to be
constant, hence single; however, additional observations may show them to be
variable, if they belong to long-period, highly eccentric systems. Figure 2
shows the $e -\log P$ diagram of our 29 orbits, which closely resembles that
of the sample of Abt \& Levy (1985). Here too, there is a lack of very short
periods ($P_{\rm orb} \la 1.0-1.5$~days): this can be understood from the fact
that no Am star has an equatorial velocity larger than 100~km\,s$^{-1}$, while
synchronism in such systems would force the primary to rotate faster. 
The distribution of periods is very similar to that of an earlier sample
(Ginestet et al. 1982), with a maximum between 5 and 10 days. But, since
observational biases tend to favour such short period systems, it remains
necessary to accumulate observations on a long timescale. On Figure 2 are
also shown the $V_r$ amplitudes $K_1$ vs. $v\sin i$ of the primaries for all 29
systems with a known orbit, as well as the $v\sin i$ of the 21 stars with an
apparently constant $V_r$. The upper envelope of the points is traced by SB2
systems with $M_2/M_1\sim 1$, 4 of which at least are already synchronized.
The mass functions of SB1 Am binaries are compatible with a main sequence
cooler secondary, as is the case of Ap binaries.
\begin{figure}[hbt]
\psfig{figure=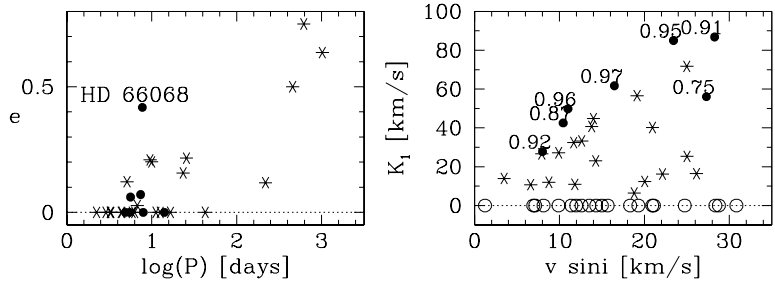,height=8cm}
\vspace{-3.5cm}
\caption{{\bf left:} $e$--$\log P$ diagram for Am stars. Key to symbols: 
asterisks=SB1 systems, full dots=SB2 systems.{\bf right:} $K_1$ vs. $v\sin i$;
same symbols as above; the open dots are for stars with a constant $V_r$.
The mass ratio is given for the SB2 systems.}
\end{figure}

\end{document}